\documentclass[12pt,onecolumn,amsmath,amssymb]{revtex4}
\begin{document}

\title{Formation of singularities on the surface of a liquid metal in a
strong electric field} 

\author{N. M. Zubarev}
\email{nick@ami.uran.ru}

\affiliation{Institute of Electrophysics, Ural Branch, Russian
Academy of Sciences,\\ 106 Amundsen Street, 620016 Ekaterinburg, Russia}

\begin{abstract}

The nonlinear dynamics of the free surface of an ideal conducting liquid
in a strong external electric field is studied.
It is establish that the equations of motion for such a liquid can be
solved in the approximation in which the surface deviates from a plane by
small angles. This makes it possible to show that on an initially smooth
surface for almost any initial conditions points with an infinite
curvature corresponding to branch points of the root type can form in a
finite time.

\end{abstract}

\maketitle

\section{INTRODUCTION}

A flat surface of a conducting liquid placed in a strong external electric
field is known \cite{1,2,3} to become unstable if the field strength $E$ exceeds
a critical value ${E_c}^2=8\pi\sqrt{g\alpha\rho}$,
where $g$ is the acceleration of free fall, $\alpha$ is the surface
tension, and $\rho$ is the density of the medium. The interaction of the
electric field and the charges induced by this field on the surface of the
liquid causes surface perturbations to grow rapidly and regions of
substantial curvature to form in a finite time \cite{4,5}. The result may be an
increase in the field energy density at the surface, initialization of
emission processes, and, finally, vacuum breakdown \cite{6}. Moreover, there
are indication that the liquid phase plays an important role in the
initial stages of explosive electron emission \cite{7}. All this means that one
must build a meaningful theoretical model of the nonlinear stages in the
development of an instability, a model that can be used to describe the
formation of a singular profile of the surface of the medium (a
liquid metal in applications). 

The present paper studies the nonlinear dynamics of an electrohydrodynamic
instability in the limit of a strong electric field, $E\gg E_c$, when both
surface tension and gravity can be ignored. The interest in this limit is
due, in particular, to the recent discovery of systems with anomalously
low critical fields, $E_c\sim 1\,\mbox{kV}\,\mbox{cm}^{-1}$
(Ref.~\cite{8}). The nonlinear stages in the development of an instability
are studied by perturbation 
techniques that use series expansions in a small parameter, the angle of
the slope of the surface. Of course, the introduction of such a small
parameter makes it impossible to describe the formation (observed in
experiments) of strong singularities, with corresponding slope angles of
order unity. Nevertheless, using the model adopted in this paper, we can
show that for almost any initial conditions at the surface of the
conducting liquid, it takes only a finite time for points with infinite
curvature to form on the surface. Thus, even in the weakly nonlinear
stages in the development of a nonlinearity there is the tendency for a
singular profile of the liquid surface to form.

The plan of the paper is as follows. In Sec.~II we derive the main
equations of motion, which describe the evolution of the free surface of
an ideal conducting liquid in a strong external electric field. In Sec.~III
we use the approximation of small angles characterizing the slope of the
surface to build a nonlinear model of the development of an
electrohydrodynamic instability. Section IV is devoted to a study of the
dynamics of one-dimensional surface perturbations. Integration of the
model equations shows that it takes only a finite time for weak
singularities of the root type to form in the system, i.e., singular
points at which the curvature of surface is infinite (see also the Letter
\cite{pla}). 

Mathematically, the formation of singularities can be explained by the
violation of the analyticity of the complex velocity potential due to the
movement of singularities, or branch points, to the boundary. On the
whole, such behavior is similar to that of the velocity potential of an
ideal liquid in the absence of external forces \cite{9,10,11}. In Sec.~V
we use the example of the evolution of single perturbations to show that the
formation of singularities occurs before the small-angle condition is
violated because of the development of a linear instability (the branch
point of the root type agrees with the small-angle approximation). In
Sec.~VI we study the behavior of the boundary of a liquid metal under the
assumption that self-similarity is retained in a small neighborhood of a
singularity in the crossover from one-dimensional perturbations of the
surface to arbitrary perturbations. Finally, in Sec.~VII we discuss the role
that branch points of the root type play in the evolution of the system.

\section{INITIAL EQUATIONS}

Consider the motion of a conducting liquid that occupies the region 
$-\infty<z\leq\eta(x,y,t)$ and is subjected to a strong electric field
$E$. We assume that this liquid is ideal and its motion is vortex-free.
Then the potential $\Phi$ of the liquid velocity is determined by the
time-dependent Bernoulli equation
$$
\Phi_t+\frac{(\nabla\Phi)^2}{2}+p/\rho=F(t),
$$
where $p$ is the pressure and $F$ is a function of time. Moreover, for
potential flow of an incompressible fluid we have $\Delta\Phi=0$.
The equations of motion must be augmented by the kinematic condition at the
free surface,
$$
\eta_t=\left.\left[\Phi_z-\nabla\eta\cdot\nabla\Phi\right]\right|_{z=\eta},
$$
by the condition at infinity, 
$\left.\nabla\Phi\right|_{z\to-\infty}\to 0$,
and the condition imposed on the pressure at the conductor-vacuum boundary
\cite{3}, 
$$
\left.\left[8\pi p+(\nabla\varphi)^2\right]\right|_{z=\eta}=0,
$$
where $\varphi$ is the potential of the electric field.

The electric potential in the absence of space charges is described by the
Laplace equation $\Delta\varphi=0$ together with the conditions that
everywhere on the surface of the conductor the potential be the same, 
$\left.\varphi\right|_{z=\eta}=0$, and that the field be uniform at
infinity, $\left.\varphi\right|_{z\to\infty}\to-Ez$.

Note that these equations of motion have a Hamiltonian structure and the
functions $\eta(x,y,t)$ and $\psi(x,y,t)=\Phi|_{z=\eta}$ are canonically
conjugate \cite{12,13}: 
$$
\frac{\partial\psi}{\partial t}=-\frac{\delta H}{\delta\eta},
\qquad
\frac{\partial\eta}{\partial t}=\frac{\delta H}{\delta\psi},
$$
where the Hamiltonian 
$$
H=\int\limits_{z\leq\eta}\frac{(\nabla\Phi)^2}{2} d^3 r
-\int\limits_{z\geq\eta}\frac{(\nabla\varphi)^2}{8\pi\rho} d^3 r
$$
coincides, to within a constant, with the total energy of the system.

\section{THE SMALL-ANGLE APPROXIMATION}

Using Green's formulas, we can write the Hamiltonian in the form of the
surface integral:
$$
H=\int\limits_{s}\left[\frac{\psi}{2}\,\frac{\partial\Phi}{\partial n}+
\frac{E^2\eta}{8\pi\rho}\,\frac{\partial\tilde\varphi}{\partial n}\right]ds,
$$
where $\tilde\varphi=(\varphi+Ez)/E$ is the perturbation of the scalar
potential, $ds$ is the surface area element, and $\partial/\partial n$ is
the normal derivative at the surface $s$. 

From now on we assume $|\nabla\eta|\ll 1$, which corresponds to
small surface-slope angles. This allows expanding the normal derivatives
in powers series of the canonical variables. Then for the Hamiltonian we
have 
$$
H=\!\int\!\frac{\psi}{2}\left(\hat T_+\hat k\hat T_+^{-1}\psi-
\nabla\eta\cdot\hat T_+\nabla\hat T_+^{-1}\psi\right) d^2 r
$$
$$
-\!\int\!\frac{E^2\eta}{8\pi\rho}\left(\hat T_-\hat k\hat T_-^{-1}\eta+
\nabla\eta\cdot\hat T_-\nabla\hat T_-^{-1}\eta\right) d^2 r.
$$
Here $\hat k$ is the two-dimensional integral operator with a kernel whose
Fourier transform is equal to the absolute value of the wave vector:
$$
\hat{k}f=-\frac{1}{2\pi}\!\int\limits_{-\infty}^{+\infty}
\int\limits_{-\infty}^{+\infty}
\frac{f(x',y')\,dx'dy'}{\left[(x'-x)^2+(y'-y)^2\right]^{3/2}}.
$$ 
The nonlinear operators $\hat T_\pm$ defined as
$$
\hat T_{\pm}=\sum_{n=0}^{\infty}\frac{(\pm\eta)^n\hat k^n}{n!}
$$
act as shift operators (i.e., $f|_{z=\eta}=\hat T f|_{z=0}$) for harmonic
functions that decay as $z\to\mp\infty$.  

If we limit ourselves to second- or third-order terms and introduce
scaling 
$$
t\to t E^{-1}(4\pi\rho)^{1/2},
\qquad
\psi\to\psi E/(4\pi\rho)^{1/2},
\qquad
H\to HE^2/(4\pi\rho),
$$
we arrive at an expression for the Hamiltonian:
\begin{equation}
H=\frac{1}{2}\int\left[\psi\hat k\psi-\eta\hat k\eta+
\eta\left((\nabla\psi)^2-(\hat k\psi)^2+(\nabla\eta)^2-
(\hat k\eta)^2\right)\right] d^2 r.
\end{equation}
The equations of motion corresponding to this Hamiltonian are
\begin{equation}
g_t+\hat k g=\frac{1}{2}
\left[(\hat k f)^2-(\nabla f)^2+(\hat k g)^2-(\nabla g)^2\right]+
\hat k\left[(f-g)\hat k f\right]+\nabla\cdot\left[(f-g)\nabla f\right],
\end{equation}
\begin{equation}
f_t-\hat k f=\frac{1}{2}
\left[(\hat k f)^2-(\nabla f)^2+(\hat k g)^2-(\nabla k g)^2\right]+
\hat k\left[(g-f)\hat k g\right]+\nabla\cdot\left[(g-f)\nabla g\right],
\end{equation}
where we have changed from the variables $\eta$ and $\psi$ to the normal
variables $f$ and $g$: 
$$
f=\frac{\psi+\eta}{2},
\qquad
g=\frac{\psi-\eta}{2}.
$$

In the linear approximation, Eq. (2) describes the relaxation of $g$ to
zero with a characteristic times $1/|k|$. In this case in the right-hand
sides of Eqs. (2) and (3) we can put $g=0$, which means we are examining
the perturbation-buildup branch with allowance for a quadratic
nonlinearity. This leads us to the following system of equations: 
\begin{equation}
g_t+\hat k g=\frac{1}{2}\,(\hat k f)^2-\frac{1}{2}\,(\nabla f)^2+
\hat k(f\hat k f)+\nabla\cdot(f\nabla f),
\end{equation}
\begin{equation}
f_t-\hat k f=\frac{1}{2}\,(\hat k f)^2-\frac{1}{2}\,(\nabla f)^2.
\end{equation}

Thus, we have shown that studying the dynamics of perturbations of the
surface of a conducting medium in a strong electric field in the
small-angle approximation amounts to studying the system of equations (4)
and (5). What is important about this system is that the nonlinear equation
(5) does not contain the function $g$ and that Eq. (4) is linear in $g$
and can easily be solved: 
\begin{equation}
g=\frac{1}{2\pi}\!\int\limits_{0}^{t}
\int\limits_{-\infty}^{+\infty}
\int\limits_{-\infty}^{+\infty}
\frac{G(x',y',t')\,(t-t')\,dx'dy'dt'}
{\left[(x'-x)^2+(y'-y)^2+(t'-t)^2\right]^{3/2}},
\end{equation}
\begin{equation}
G(x,y,t)=\frac{1}{2}\,(\hat k f)^2-\frac{1}{2}\,(\nabla f)^2+
\hat k(f\hat k f)+\nabla\cdot(f\nabla f),
\end{equation}
where we assumed that $g|_{t=0}=0$.

\section{FORMATION OF A BRANCH POINT OF THE ROOT TYPE}

In the case of one-dimensional perturbations of the surface (we ignore the
dependence of all the quantities on $y$), the integral operator $\hat k$
can be expressed in terms of the Hilbert operator $\hat{H}$:
$$
\hat k=-\frac{\partial}{\partial x}\,\hat H,
\qquad
\hat{H}f=\frac{1}{\pi}\!\!\int\limits_{-\infty}^{+\infty}
\frac{f(x')}{x'-x}\,dx'.
$$
Then the model equations (4) and (5) can written
\begin{equation}
g_t-\hat H g_x=\frac{1}{2}\,(\hat H f_x)^2-\frac{1}{2}\,(f_x)^2+
\hat{H}(f\hat{H}f_x)_x+\left(ff_x\right)_x,
\end{equation}
\begin{equation}
f_t+\hat H f_x=\frac{1}{2}\,(\hat H f_x)^2-\frac{1}{2}\,(f_x)^2.
\end{equation}

For further discussions it is convenient to introduce functions that are
analytic in the upper half-plane of the complex variable $x$:
$$
\phi=\hat{P}f, \qquad \chi=\hat{P}g,
$$
where $\hat{P}=(1-i\hat H)/2$. Since applying the Hilbert operator to a
function that is analytic in the upper half-plane amounts to multiplying
that function by the unit imaginary number, Eqs. 
(8) and (9) take the form 
$$
\mbox{Re}\left(\phi_t+i\phi_x+\phi_x^2\right)=0,
$$
$$
\mbox{Re}\left(\chi_t-i\chi_x+\phi_x^2
-2\hat{P}\left(\phi\bar{\phi}_x\right)_x\right)=0.
$$
Thus, the integro-differential equations (8) and (9) can be studied simply
by analyzing the inhomogeneous linear equation
\begin{equation}
\chi_t-i\chi_x=-\phi_x^2+2\hat{P}\left(\phi\bar{\phi}_x\right)_x
\end{equation}
and (separately) the nonlinear partial differential equation 
\begin{equation}
\phi_t+i\phi_x=-\phi_x^2.
\end{equation}

For the sake of convenience we introduce a new function, $v=\phi_x$. In
terms of this function, Eq. (11) becomes 
$$
v_t+iv_x=-2vv_x.
$$
Note that this equation coincides with the one proposed by Zhdanov and
Trubnikov \cite{14,15}, who used it to describe the nonlinear stages in the
development of tangential discontinuities in hydrodynamics. More than
that, if we replace $x$ by $x\to x+it$, we arrive at the equation derived
in Refs.~\cite{9,10,11} as a result of a discussion of the nonlinear dynamics of
a free surface of ideal liquid in the absence of external forces, where it
describes the temporal evolution of the complex-valued velocity. The
solution of this first-order partial differential equation can be found by
using the method of characteristics: 
\begin{equation}
v=Q(x'),
\end{equation}
\begin{equation}
x=x'+it+2Q(x')t,
\end{equation}
where the function $Q$ is determined by the initial conditions
$Q(x)=v|_{t=0}$. 

Let us show, by analogy with Refs. \cite{9,10,11}, that these relations describe
(if we require that $v$ be analytic) the formation of a singularity in a
finite time. The problem of finding the explicit solution reduces to
analyzing the map $x\to x'$ specified by Eq. (13). Generally, this map
ceases to be one-to-one at points where
\begin{equation}
\partial x/\partial x'=1+2Q_{x'}t=0.
\end{equation}
The relationship (14) specifies a path $x'=x'(t)$ in the complex $x'$
plane. Then the motion of the branch point of the function $v$ is given by
$$
x(t)=x'(t)+it+2Q(x'(t))t.
$$
At the time $t_0$ when the branch point reaches the real axis the
analyticity of $v$ is violated and the solutions of Eq. (9) become singular.

Let us examine the behavior of the solutions near a singularity. Expanding
(12) and (13) in a small neighborhood of the point $t=t_0$,
$x=x_0=x(t_0)$, $x'=x'_0=x'(t_0)$, in the leading order we get
$$
v=q_0-\delta x'/(2t_0),
\qquad
\delta x=i\delta t+2q_0\delta t+q''t_0(\delta x')^2,
$$
where $q_0=Q(x'_0)$, $q''=Q_{x'x'}(x'_0)$, 
$\delta x=x-x_0$, 
$\delta x'=x'-x'_0$, and
$\delta t=t-t_0$.

Excluding $\delta x'$ from these expressions, we obtain
\begin{equation}
v=q_0-\left[\frac{\delta x-i\delta t-2q_0\delta t}
{4q''t_0^3}\right]^{1/2}.
\end{equation}
The derivative of this expressions with respect to $x$ is
\begin{equation}
\phi_{xx}\equiv v_x=-\left[16q''t_0^3
(\delta x-i\delta t-2q_0\delta t)\right]^{-1/2},
\end{equation}
which shows that $\phi_{xx}$ behaves in a self-similar manner ($\delta
x\sim\delta t$) and becomes infinite as $\delta t\to 0$. 

As for the complex-valued function $\chi$, the equation that determines
its temporal dynamics (Eq.~(10)) can be integrated by the method of
characteristics (see Eqs. (6) and (7)). Taking the initial condition in
the form $\chi|_{t=0}=0$ yields 
$$
\chi=\int\limits_{0}^{t} D(x+it-it',t')\,dt', \qquad
D(x,t)=-\phi_x^2+2\hat{P}\left(\phi\bar{\phi}_x\right)_x.
$$
Inserting (15) into this expression, we see than near the singularity the
derivative $\chi_{xx}$ can be expressed in terms of $\phi_{xx}$: 
$$
\chi_{xx}=\frac{(\bar{q_0}-q_0)}{(q_0+i)}\,\phi_{xx}.
$$
This means that the analyticity of $\chi_{xx}$ is violated at time $t=t_0$.

How does the surface of the liquid metal behave at the time when the
singularities develop in the solutions of Eqs. (10) and (11)? Allowing for
the fact that $\eta=f-g$, we find that the surface curvature  
$$
K=\eta_{xx}\left(1+\eta_x^2\right)^{-3/2}
$$
is specified, to within a quadratic nonlinearity, by the expression 
$$
K\approx\eta_{xx}=2\mbox{Re}\,(\phi_{xx}-\chi_{xx}).
$$
Substituting the expression for $\phi_{xx}$ and $\chi_{xx}$ found earlier,
we find that in a small neighborhood of the singular point
\begin{equation}
K\approx 2\mbox{Re}\left[1-\frac{(\bar{q_0}-q_0)}{(q_0+i)}\right]
\phi_{xx}.
\end{equation}
Since $\phi_{xx}$ is given by (16), we have  
$$
K|_{x=x_0}\sim |\delta t|^{-1/2}, \qquad
K|_{t=t_0}\sim |\delta x|^{-1/2},
$$
i.e., it takes a finite time for a singularity of root type (branch point)
to form at the surface, and the curvature of the surface of the liquid at
this point is infinite.

To conclude this section we note that since we have $\psi=f+g$, near the
singularity a relationship holds for the complex-valued potential of the
liquid flow, $\Psi\equiv 2\hat P\psi$: 
$$
\Psi_{xx}=2(\phi_{xx}+\chi_{xx})
\approx 2\frac{(\bar{q_0}+i)}{(q_0+i)}\,\phi_{xx},
$$
i.e., the first derivative of the complex-valued velocity also exhibits
singular behavior as $\delta t\to 0$. This means that, as in Refs.
\cite{9,10,11}, 
the formation of singularities can be interpreted as the result of
violation of the analyticity of the complex-valued potential due to the
movement of the singularities of the potential to the boundary.

\section{EVOLUTION OF A SINGLE PERTURBATION}

We use a simple example to show that at the time when a singularity in the
solutions of Eqs. (8) and (9) develops the applicability conditions for
our model are met.

We take the initial perturbation in the form
\begin{equation}
f|_{t=0}=-\varepsilon a^m\hat{k}^{m-1}\ln(x^2+a^2),
\end{equation}
where $m$ is a positive integer, and the parameters $a$ and $\varepsilon$
take positive values $a>0$ and $\varepsilon>0$. This situation corresponds
to a one-dimensional single perturbation of the surface symmetric with
respect to point $x=0$, at which the surface curvature is negative. The
characteristic slope angles of the surface are determined by the parameter
$\varepsilon$, which we assume small. 

Note that in the linear approximation Eq. (9) becomes
$$
f_t+\hat{H}f_x=0.
$$
Its solution with the initial condition (18) is
$$
f(x,t)=-\varepsilon a^m\hat{k}^{m-1}\ln\,(x^2+(a-t)^2),
$$
i.e., within the linear model the perturbation grows without limit and
becomes infinite at the time $t=a$, which of course violates the
applicability conditions for this model.

Will introducing nonlinearity into the model permit a singularity to
develop in the solution before the condition $|\eta_x|\approx|f_x|\ll 1$
breaks down? (The branch-point nature of this singularity agrees with the
condition that the angles be small.) To answer this question, we will
examine the evolution of the perturbation (18) according to the nonlinear
equation (9). 

The symmetry of (18) implies that the singularity forms at point $x=0$.
Then from (13) and (14) it follows that the time $t_0$ at which the
singularity develops can be found by solving the following equations
simultaneously: 
$$
x_0'+it_0+2Q(x'_0)\,t_0=0,
\qquad
1+2Q_{x'}(x'_0)\,t_0=0,
$$
where the function $Q$ corresponding to (18) has a pole of order $m$ at
the point $x'=-ia$: 
$$
Q(x')=i\varepsilon(m-1)!\left(\frac{ia}{x'+ia}\right)^m.
$$
Expanding in a power series in the small parameter $\varepsilon$, we
obtain to leading order the following:
$$
t_0\approx a\left[1-\frac{m+1}{m}\,
\left(2\varepsilon m!\right)^{1/(m+1)}\right],
\qquad
x'_0\approx-ia\left[1-\left(2\varepsilon m!\right)^{1/(m+1)}\right].
$$
Since in the linear approximation the singularity is formed at time $t=a$,
the above expression for $t_0$ implies that the nonlinearity accelerates
the formation of the singularity (but if $\varepsilon<0$ holds, the 
nonlinearity delays the onset of the instability).

Plugging the above expression for $x'_0$ into the expression for $Q$ and
its second derivative $Q_{x'x'}$, we obtain
$$
q''\approx-\frac{i(m+1)}{2a^2}\left(2\varepsilon m!\right)^{-1/(m+1)},
\qquad
q_0\approx\frac{i}{2m}\left(2\varepsilon m!\right)^{1/(m+1)}.
$$
Thus, for perturbations of the form (18) the parameter $q''$ is finite. 
This means that the dynamics of surface perturbations near a singular
point is described fairly accurately by Eqs. (15)--(17). As for the
parameter $q_0$, the smallness of $\varepsilon$ implies $|q_0|\ll 1$. This
is an important result. The important point is that this parameter, as
(12) and (15) imply, determines the characteristic angles of slope of the
surface by the moment of singularity formation. Then for the derivative 
$\eta_x$ at the time of collapse the following estimate holds: 
$$
|\eta_x|\sim\varepsilon^{1/(m+1)}\ll 1,
$$
i.e., the characteristic angles remain small, even through they increased
by a factor of $\varepsilon^{-m/(m+1)}$. In this case there is not enough
time for the small-angle condition to be violated as a result of the
development of a linear instability, and the proposed model (Eqs. (8) and
(9)) is closed in the sense that if the initial perturbation meets all the
conditions needed for the model to be valid, this property is retained
throughout the entire evolution until the time of collapse, $t_0$.  

We now discuss the behavior of a perturbation of the electric field at the
conducting surface,
$$
\delta E(x,t)\equiv-E-
\left.\frac{\partial\varphi}{\partial n}\right|_{z=\eta}
$$
near the singularity. Clearly, in the linear approximation the field
perturbation is specified by the formula
$$
\delta E\approx-E\,\hat{H}f_x=2E\,\mbox{Im}\,(v).
$$
Substituting $v$ of Eq. (15) in this expression, we find that at the
singular point 
$$
\delta E|_{\delta x=\delta t=0}\approx 2E\,\mbox{Im}\,(q_0).
$$
Since the parameter $q_0$ is small, the perturbation of the electric field
at the time of singularity formation remain much weaker than the external
field (both $\delta E_x$ and $\delta E_t$ are singular).

\section{SELF-SIMILAR SOLUTIONS IN THE GENERAL CASE}

In Sec.~5 we found that $|q_0|\ll 1$ holds in the small-angle approximation.
This allows us to ignore the dependence on this parameter in Eqs.~(16) and
(17). Then, if $q''\not= 0$, the dynamics of the surface curvature in the
one-dimensional case is given by the expression
\begin{equation}
K\approx-\mbox{Re}\left[4q''t_0^3
(\delta x-i\delta t)\right]^{-1/2}.
\end{equation}
In particular, for a perturbation that is symmetric with about the point
$x=x_0$ we have 
$$
K\approx-\frac{1}{\sqrt{8t_0^3|q''|}}
\left[\frac{-\delta t+\sqrt{\delta x^2+\delta t^2}}
{\delta x^2+\delta t^2}\right]^{1/2}.
$$
Here the initial conditions determine only the constant factor. Thus, the
behavior of the system near a singular point is of a universal nature.

Let us take a particular case $q''=0$. Suppose, for instance, that
$$
\left.\frac{\partial^2 Q}
{\partial {x'}^2}\right|_{x'=x'_0}=...
=\left.\frac{\partial^{n-1} Q}
{\partial {x'}^{n-1}}\right|_{x'=x'_0}=0,
\qquad\qquad
q_n\equiv\left.\frac{\partial^n Q}
{\partial {x'}^n}\right|_{x'=x'_0}\not=0,
$$
where $n>2$. Then, expanding (13) in a power series in $\delta x'$ up to
the $n$th power, in the leading order we get
\begin{equation}
K\approx-\frac{1}{nt_0}\,\mbox{Re}
\left(\frac{n!}{2t_0q_n}\right)^{1/n}\!\!
\left(\delta x-i\delta t\right)^{1/n-1}. 
\end{equation}

The formulas (19) and (20) show that for an arbitrary one-dimensional
perturbation of the surface satisfying the condition $|\eta_x|\ll 1$ 
the curvature near
the singularity behaves self-similarly:
\begin{equation}
K\approx\frac{1}{|\delta t|^p}\,h\left(\frac{\delta x}{|\delta t|}\right),
\end{equation}
where $h$ is an unknown function, and the exponent is given by the
expression 
\begin{equation}
p=(n-1)/n.
\end{equation}
with $n$ a positive integer.

Note that (19) and (20) are the exact solutions of the linear
integro-differential equation
$$
K_t+\hat H{K}_x=0,
$$
which describes the evolution of the surface curvature in the linear
approximation, as follows from (9) with allowance for the fact that
$K=f_{xx}$ holds in the leading order. For an arbitrary exponent $p$, Eq.
(21) specifies the class of self-similar solutions of this equation. This
means that, on the one hand, the dynamics of the surface near a
singularity is described by the self-similar solutions of the linearized
equations of the model and, on the other hand, that the presence of a
nonlinearity leads to a situation in which of all the possible
self-similar solutions only those with rational values of $p$ specified by
the condition (22) are realized (from general considerations it follows
that $p=1/2$). 

It is therefore natural to assume that in the two-dimensional case, as in
the one-dimensional, the solutions in a small neighborhood of the
singularity are self-similar: 
\begin{equation}
K\approx\frac{1}{|\delta t|^p}\,
h\left(\frac{\delta x}{|\delta t|},\frac{\delta y}{|\delta t|}\right),
\end{equation}
where $p$ satisfies the condition (22). A characteristic feature of the
weak-nonlinearity approximation in our problem is that the specific form
of the dependence of all quantities on the self-similar variables can be
treated using the equation 
\begin{equation}
K_t=\hat{k}K, 
\end{equation} 
whose linearity makes it possible to effectively study the formation of
two-dimensional singularities at the surface of a conducting medium. Note
that this statement is valid if $p<1$, which, as condition (22) shows, is
met in our case in a natural manner. The point is that at $p=1$ an
expression of the form (23) corresponds to the symmetries of the initial
nonlinear equation of motion. This mean that near a singularity the
contribution of a nonlinearity is comparable to that of the linear terms,
and the analysis of the behavior of the surface lies outside the scope of
this paper.

Substituting (23) in (24), we arrive at the following integro-differential
equation for the unknown function $h$:
$$
\xi h_{\xi}+\zeta h_{\zeta}+ph=\hat{k}(\xi,\zeta)h,
$$
where $\xi=\delta x/|\delta t|$ and $\zeta=\delta y/|\delta t|$.
Since the profile of the surface begin to form at the periphery and only
then is propagated to the point $\delta x=\delta y=0$, at the time of
collapse the curvature of the surface in a small neighborhood of the
singular point is determined by the asymptotic solutions of this equation
as $\xi^2+\zeta^2\to\infty$. As can easily be shown, these solutions are
described by the partial differential equation
$$
\xi h_{\xi}+\zeta h_{\zeta}+ph=0,
$$
whose general equation is
$$
h=\left[\xi^2+\zeta^2\right]^{-p/2}F(\zeta/\xi),
$$
where $F$ is an unknown function. Plugging this expression into (23) and
introducing polar coordinates, 
$$
\delta x=r\cos{\beta},
\qquad
\delta y=r\sin{\beta},
$$ 
we arrive at the following formula for the curvature of the surface near
the singular point: 
$$
K|_{t=t_0}\approx\frac{F\left(\mbox{tg}\,\beta\right)}{r^p}.
$$
We see that we are again dealing with a branch point of the root type.

\section{CONCLUSION}

Our analysis of the evolution of the boundary of a conducting liquid in a
strong electric field within the small-angle approximation has shown that
for almost any initial conditions on an initially smooth surface the
presence of a nonlinearity gives rise to points at which the curvature of
the surface becomes infinite. These points correspond to branch points of
the root type. However, the presence of such singularities does not ensure
a significant concentration of the electric field near the surface of the
conductor and, hence, cannot by itself lead to vacuum breakdown. In this
case we may assume that the main role of these branch points in the
general evolution of the system is, in time, to generate stronger
singularities capable of substantially influencing the emission from
liquid metal; in particular, capable of ensuring the conditions needing
for the initiation of explosive electron emission. Among such
singularities are, for instance, discontinuities in the first derivative
of the surface profile, which were observed in experiments \cite{5,7}.
Note that a theoretical study of such singularities lies outside the scope
of the small-angle approximation and requires allowing for surface
tension. Indeed, the applicability of the model adopted in the present
paper is limited to the scales $|\delta x|\gg l$, where the parameter $l$
is the characteristics length on which the capillary effects become
important. This parameter, on dimensional grounds, is determined by the
ratio of surface tension to the electric-field energy density:
$l\sim\alpha/E^2$. The stabilizing effect of the surface pressure means
that at time $t_0$ the curvature of the surface is still finite $(K\sim
1/l)$ and the profile of the surface is smooth, and the formation of a
singular profile begins in the later stages in the development of an
instability.

\medskip

The author would like to express his gratitude to A. M. Iskol'dskii and N. B.
Volkov for stimulating discussions and to E. A. Kuznetsov who kindly
pointed out Refs. \cite{9,10,11}.

\end {document}